# Batch effects can impair federated learning in multi-center omics studies

Yuliya Burankova[1,2,*,†], Julian Klemm[1,†], Jens J. G. Lohmann[1], Anne Hartebrodt[3,4], Ahmad Taheri[1], Niklas Probul[1], Jan Baumbach[1,3,‡], Olga Zolotareva[1,5,6,*,‡]

[1] *Institute for Computational Systems Biomedicine, University of Hamburg, 22761 Hamburg, Germany*
[2] *Chair of Proteomics and Bioanalytics, TUM School of Life Sciences, Technical University of Munich, 85354 Freising, Germany*
[3] *Department of Mathematics and Computer Science, University of Southern Denmark, 5230 Odense, Denmark*
[4] *Biomedical Network Science Lab, Department Artificial Intelligence in Biomedical Engineering, Friedrich-Alexander-Universität Erlangen-Nürnberg, 91052 Erlangen, Germany*
[5] *Data Science in Systems Biology, TUM School of Life Sciences, Technical University of Munich, 85354 Freising, Germany*
[6] Institute of Clinical Molecular Biology (IKMB), Kiel University and University Medical Center Schleswig-Holstein, Rosalind-Franklin-Str. 12, 24105 Kiel, Germany

*To whom correspondence should be addressed:
Yuliya Burankova, Email: yuliya.burankova@tum.de
Julian Klemm, Email: julian.klemm@uni-hamburg.de
Correspondence may also be addressed to Olga Zolotareva. Email: o.zolotareva@ikmb.uni-kiel.de
†The first two authors should be regarded as Joint First Authors.
‡The last two authors should be regarded as Joint Last Authors.

## Abstract

Federated learning (FL) enables collaborative analysis of biomedical data without exchanging sensitive patient-level information, but its performance in multi-center studies may be compromised by batch effects which can obscure biological signals. Here, we systematically assess the impact of uncorrected batch effects on FL outcomes using four multi-center omics datasets, including transcriptomic, proteomic, and metabolomic data, and two representative algorithms: federated k-means clustering and federated random forest classification.

Our results demonstrate that uncorrected batch effects undermine unsupervised FL and can substantially degrade supervised FL performance, indicating that privacy-aware batch-effect correction is essential for reliable FL.

To enable privacy-preserving BEC in distributed bulk omics data, we introduce fedRBE (https://featurecloud.ai/app/fedrbe), a federated implementation of limma's removeBatchEffect() method enhanced by secure multi-party computation, suitable for datasets with missing values and non-identical feature sets across clients, including proteomics and metabolomics data.

# Introduction

The rise of machine learning and artificial intelligence (AI) has revolutionized biomedical sciences, enabling the identification of novel biomarkers and the prediction of disease outcomes, helping to enhance personalized medicine and accelerating drug discovery[1,2]. Yet building effective models often requires large, high-quality biomedical datasets[3]. Since individual research centers often have limited sample sizes, multi-center collaborations become necessary for creating advanced AI models. Traditionally, such collaborations depend on centralized data aggregation, which raises privacy, legal, and ethical concerns[4], especially for sensitive data[5], such as patient health records[6], omics data[7], and clinical trial results[8]. Researchers must comply with regulations like the General Data Protection Regulation (GDPR)[9], the California Consumer Privacy Act (CCPA)[10], and the Health Insurance Portability and Accountability Act (HIPAA)[11]. To avoid unprotected data sharing and enable a privacy-aware analysis of distributed multi-center data, many computational tools that employ various privacy-enhancing techniques have been developed[12].

Despite these advances, many analyses still remain challenging in a distributed setting due to batch effects — systematic biases in measurements arising from differences in experimental conditions or equipment across data-generating sites. Batch effects are a common problem for many omics data types[13,14], including microarrays, sequencing-based transcriptomics, and mass spectrometry (MS) proteomics data. Unaccounted batch effects obscure true biological variation and can impair model performance, leading to biased and non-reproducible results[14,15]. Consequently, many data analysis tasks, including clustering[16], dimensionality reduction[17], and multi-omics integration[18], often require batch effect correction (BEC) to be applied to the data. To address this challenge, numerous methods have been developed to computationally remove batch effects from the pooled multi-center data[19–21]. This diversity reflects the distinct characteristics of different biomedical data types, including data distributions and missingness patterns, which require tailored BEC approaches.

In contrast, many works involving decentralized data analysis do not explicitly model batch effects in multicenter datasets[22,23]. Instead, they rely on benchmarks based on random splits of large single-center datasets, where batch effects are minimal or absent[24,25]. Although some studies have used real multicenter data[26–28], the effect of BEC on downstream analyses in federated settings remains insufficiently characterized.

We hypothesize that, as in centralized analyses, unaccounted batch effects in decentralized data can adversely affect downstream results. Despite the importance of this issue, only a few tools for batch effect correction in decentralized settings are currently available. FedscGen and federated Harmony were recently developed for federated single-cell RNA-seq data integration[29,30]. However, scRNA-seq data differ substantially from bulk omics data, which may limit the direct applicability of these methods. In addition, FedscGen is based on deep learning, and such methods typically require large sample sizes[31], whereas bulk omics datasets often contain only tens to hundreds of samples[32].

A potentially more suitable tool for federated BEC in bulk omics data is d-ComBat[33], an adaptation of ComBat[34] for distributed data. However, d-ComBat requires manual coordination of data exchange and has important security limitations because it exposes design matrices, making them potentially vulnerable to reconstruction attacks. Most importantly, FedscGen, federated Harmony, and d-ComBat generally require preprocessing or imputation of missing values before analysis, which limits their applicability to bulk omics datasets with extensive non-random missingness.

In this study, we empirically assess how batch effects in real multicenter datasets affect downstream federated analyses. We evaluate this effect using four real multi-center datasets and consider both supervised and unsupervised learning tasks. In addition to existing tools for federated BEC unable to handle data with missing values, we propose a privacy-preserving version of the popular BEC method, removeBatchEffect() from the *limma* R package[19]. Unlike other BEC methods for decentralized data[29,30,33], fedRBE allows inputs with missing values and is well suited for omics data with missingness, such as proteomics data[35] or metabolomics[36] data.

# Results

## Methodological design of federated BEC

We selected *limma*'s removeBatchEffect()[19] method as the foundation for the federated BEC tool because it combines several properties important for BEC of bulk omics data. Unlike ComBat and its adaptations[33,34,37], *removeBatchEffect()* can operate on datasets containing missing values, which is particularly important for proteomics[35] or metabolomics[36] applications. In addition, it enables adjustment for covariates, thereby reducing the risk of over-correction in the presence of known biological factors correlated with the batch.

To enable BEC in distributed data while maintaining patient data privacy, FedRBE utilizes a combination of federated learning[38] (FL) and secure multi-party computation[39] (SMPC). FL is a privacy-enhancing technique that allows multiple institutions to collaboratively train models without exchanging raw data. Each participant performs local computations using its data and exchanges only selected intermediate results or model parameters, while keeping original data securely stored in its local environment[38,40–43]. FL has recently demonstrated its ability to achieve performance similar to centralized analysis across multiple biomedical applications[22,23,27,28,44]. To prevent information leakage from the model parameters shared by the participants, FL can be further enhanced by combining it with SMPC[45] (**Figure 1**). In SMPC, the participants split their local data into random shares, which are further exchanged and aggregated, so that the desired computational result is reconstructed without revealing any local data.

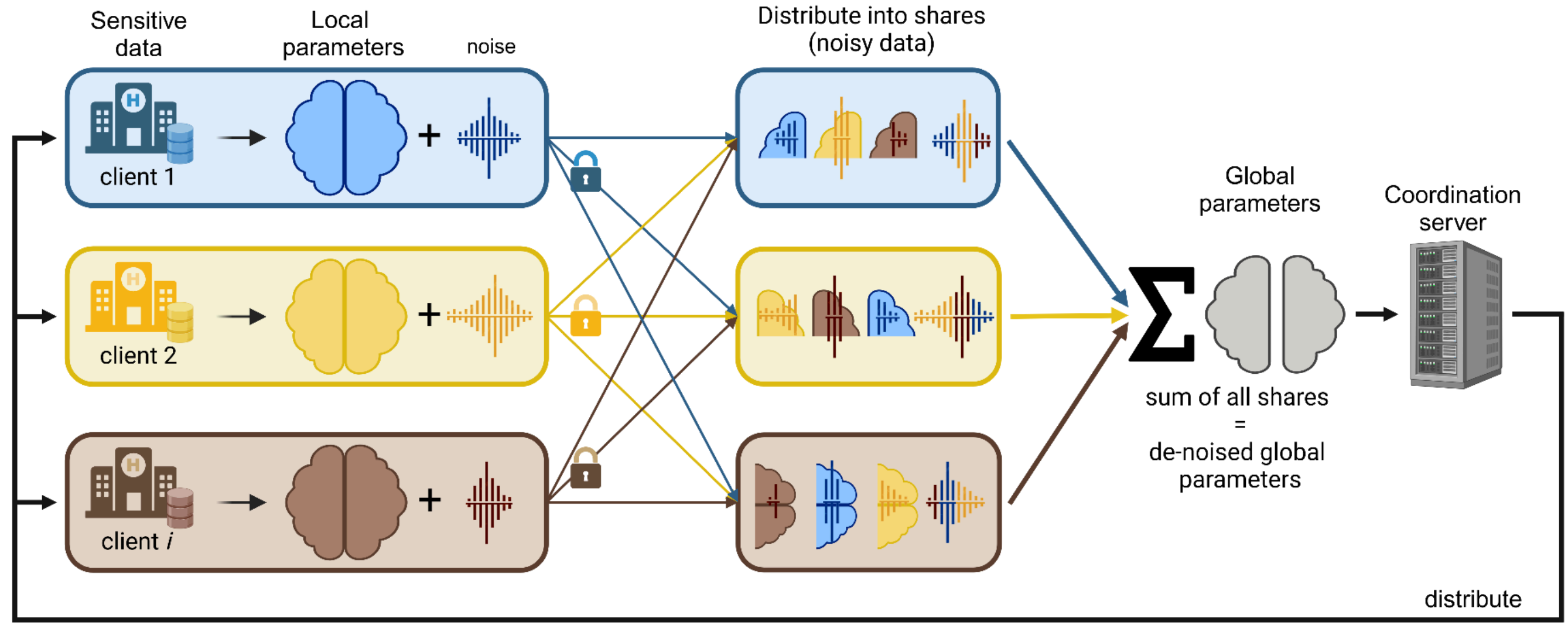

**Figure 1. A privacy-preserving workflow integrating FL and SMPC.**
In classical FL, $c$ participants collaboratively train a model on distributed data by sharing intermediate results with a coordination server which aggregates them to the model. The intermediate results are computed by each participant on their local data[38]. To avoid reconstruction attacks, FL is enhanced by SMPC via additive secret sharing[39,46], where each participant masks local data with random noise and partitions it into $c$ shares before sharing (see Matschinske et al., 2023[46] and Burankova et al., 2025[28] for details).

## BEC in distributed omics data with fedRBE

FedRBE is implemented as a FeatureCloud App[46], accessible for users through an intuitive graphical interface[47]. To initialize the workflow, the coordinating user creates a fedRBE project on the FeatureCloud website by adding the fedRBE application into the workflow (**Supplementary Figure S1**). The coordinator then specifies analysis parameters, defines the expected number of participating centers, and distributes invitation tokens to participants.

Each participant subsequently registers on FeatureCloud, launches their client applications, and joins the fedRBE project using the provided token. Each participant prepares a single flat *.zip* archive containing *data.csv* and *config.yml* and uploads it through the FeatureCloud GUI (see **Supplementary Figures S2 and S3**). In cases where there are multiple batches per client or any covariates that must be taken into account, the participants should add an additional *design.csv* file with relevant metadata and update their *config.yml* accordingly (see **Supplementary Figure S2**). The *.zip* archive is exclusively used locally by the Docker container with the fedRBE app; no raw information ever leaves the originating site or is shared with the coordinator or other clients.

Once all the invited participants have joined the project, fedRBE performs federated computations as described in Figure 2. The coordinating user's client acts as the coordinating server as well as an individual client. First, the coordinating server finds the intersection of covariates, guaranteeing consistent beta coefficient dimensions, and the union of features, and broadcasts both to all participating sites. The union of features ensures that as many features as possible are being batch effect corrected, e.g., a feature available in three clients but not in all five clients can still be corrected.

Further, each client computes the intermediate matrices $X_i^T X_i$ and $X_i^T y_i$, which are further aggregated to global $X^T X$ and $X^T y$ via the SMPC protocol[46] and used by the coordinator to calculate the global beta coefficients. Finally, each client receives these coefficients from the coordinator and applies them to correct batch effects in their local data. The fedRBE algorithm, explained in detail in the next section, yields results that are effectively identical to the original removeBatchEffect() method applied to the same data collected centrally and pooled.

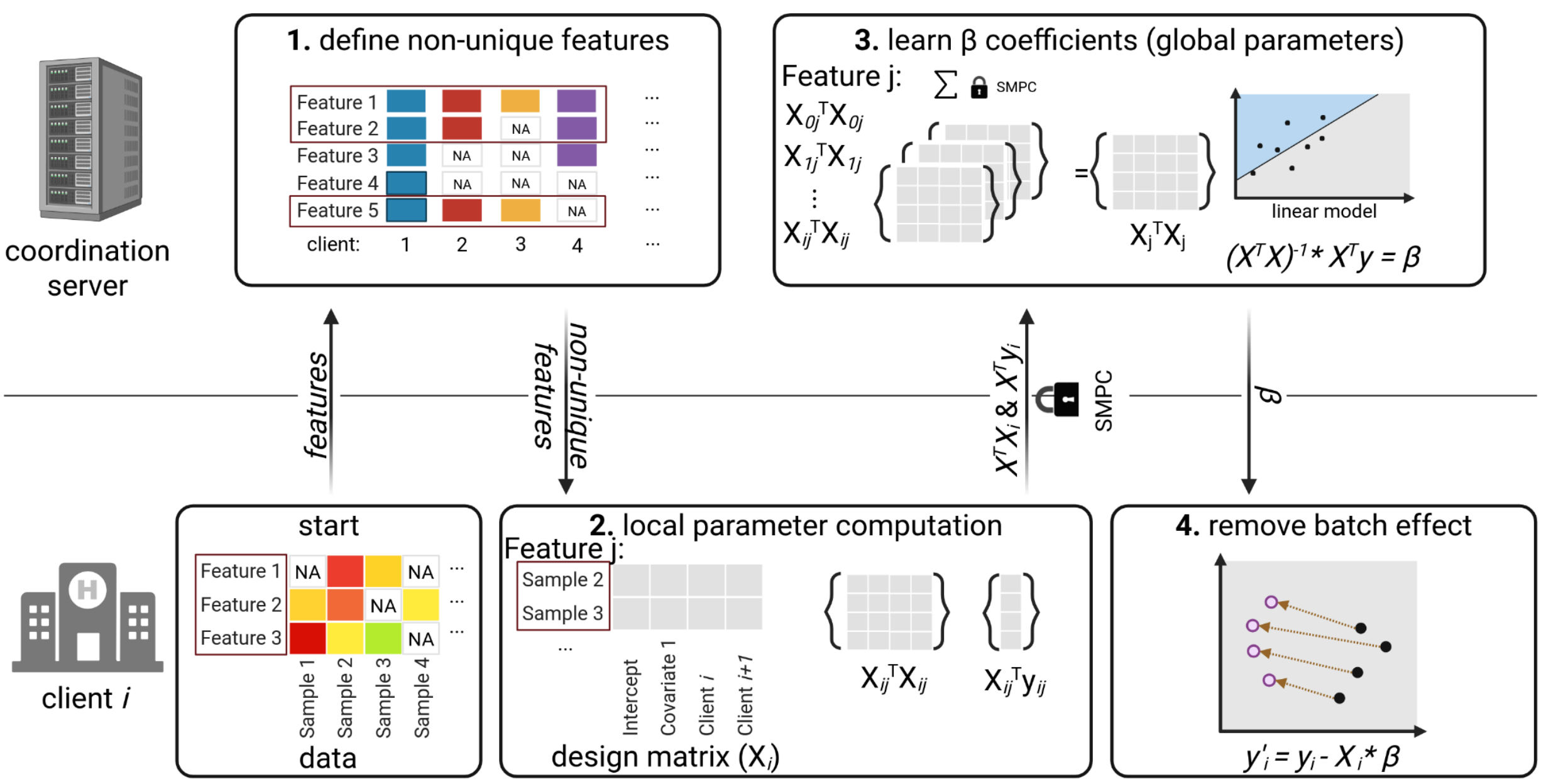

**Figure 2. The fedRBE workflow overview.**
Federated computation begins once all clients have joined and received the configuration file along with two CSV files containing the local data matrix $y_i$ and the design matrix $X_i$. In the first step, the coordinator collects the feature names from the clients and identifies the shared feature names present in at least three of the clients. Only these features are passed to further analysis, since SMPC requires data from at least three parties. In the second step, the clients calculate their local $X_i^T X_i$ and $X_i^T y_i$ and employ SMPC to aggregate them to global results. This ensures that the coordinator receives only the aggregated global $X^T X$ and $X^T y$ but not individual $X_i^T X_i$ or $X_i^T y_i$, which theoretically could be used to reconstruct a part of the original data in some cases. Finally, the coordinator calculates the global matrix of regression coefficients β and returns it to the clients, which use it to regress out batch effects from their local data.

## FedRBE yields the same results as centralized BEC

Firstly, to assess the quality of fedRBE results, we applied it to four multi-center datasets of different omics types and compared its outputs numerically to the outputs of *limma*'s removeBatchEffect() applied to the same data, collected centrally and pooled. Each dataset consists of several cohorts measured at distinct locations, inherently introducing batch effects. Specifically, we used four datasets: (i) a publicly available multi-center mass spectrometry-based proteomics dataset consisting of 118 *Escherichia coli* samples from five laboratories[28]; (ii) microarray-based gene expression profiles of ovarian tissue samples from patients with ovarian cancer and from healthy controls from six independent studies[48–53], (iii) clear cell renal cell carcinoma (ccRCC) proteomics data from three laboratories[54–56], and (iv) the Quartet dataset — transcriptomic, proteomic, and metabolomic profiles of lymphoblastoid cells obtained from a pair of twins and their parents[57]. Detailed descriptions of the datasets are provided in **Table 1** and **Supplementary Table S1**.

In all four datasets, sample clustering before batch effect correction was dominated by cohort and batch effects, masking biological differences (**Figure 3**, **Supplementary Figure S5**). After FedRBE correction, batch-specific variance fell from 15–99% before correction to essentially zero in all cases after BEC. We compared fedRBE results with those from the removeBatchEffect() method applied to the same data collected centrally and pooled using the union of features. For all multi-center datasets, fedRBE reproduced *limma's* removeBatchEffect() results with negligible differences (max absolute element-wise difference $\leq 2.49\times10^{-13}$; **Supplementary Table S2**).

In addition, to examine the effect of data imbalance on the results of fedRBE, we generated artificial datasets simulating three scenarios with increasing levels of imbalance, following the approach proposed by Burankova et al. 2025[28] (see Methods). For each scenario, data generation was repeated 30 times, and the difference between the results of fedRBE and *limma*'s removeBatchEffect() was calculated. We used the same number of samples, batch distribution, and scenarios (balanced, mildly, and strongly imbalanced data; see **Supplementary Table S3**). On the simulated data, fedRBE results closely matched *limma*'s outputs as well, regardless of the degree of imbalance in the data (mean absolute differences $\leq 3\times10^{-15}$, maximum absolute difference $\leq 5\times10^{-13}$; **Supplementary Table S4** and **Figure S4**).

**Table 1. Composition of multi-center datasets used for fedRBE evaluation.**
Target classes represent groups of interest in the data, the difference between which is being studied, and should be accounted for during BEC. *- For the multi-omics Quartet dataset, a subset of 48 transcriptomic, proteomic, and metabolomic profiles from the Quartet dataset[57] was used to simulate three clients (see Methods and **Supplementary Table S1** for details).

| **Multi-center dataset** | **Batches** | **Features** per center | **Samples** per center | **Target classes and Samples** per target class |
|---|---|---|---|---|
| MS-based proteomics *E. coli* dataset[28] | 5 centers | 2401 – 2846 | 23 – 24 | glucose (59), pyruvate (59) |
| Microarray-based ovarian cancer transcriptomics | GSE6008[48] | 21128 – 51276 | 19 – 195 | tumors (279), controls (53) |
| | GSE26712[49] | | | |
| | GSE40595[50] | | | |
| | GSE69428[51] | | | |
| | GSE38666[52] | | | |
| | GSE14407[53] | | | |
| ccRCC proteomics | PDC000127[54] | 7655 – 12548 | 194 – 464 | tumors (457), controls (430) |
| | PXD030344[55] | | | |
| | PXD042844[56] | | | |
| Quartet multi-omics[57] | 3 clients* | 26907 gene expressions, 2539 protein expressions, 71 metabolite abundances | 12 – 24 | D5 (12), D6 (12), F7 (12), M8 (12) |

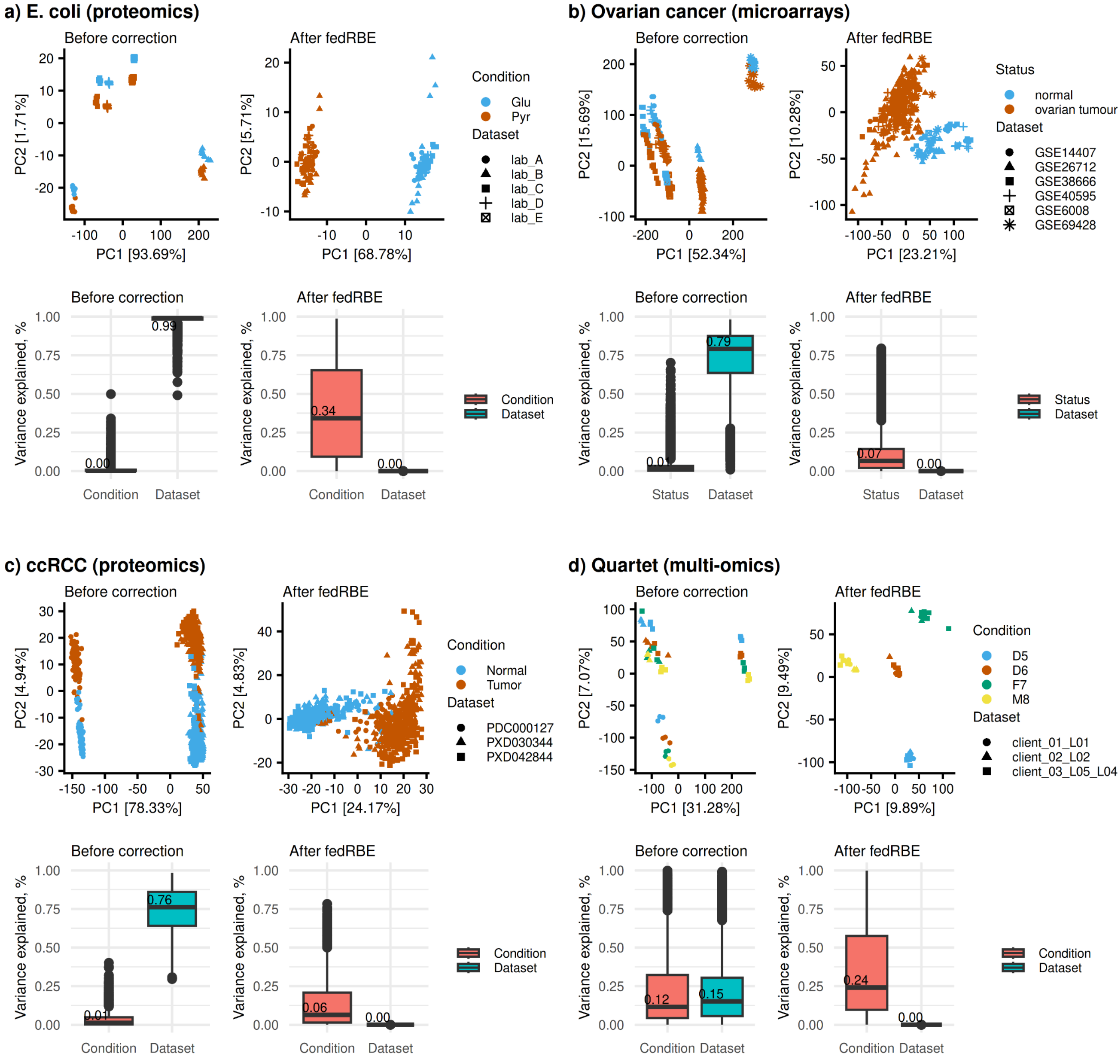

**Figure 3. The results of batch effect correction of real-world datasets using fedRBE.**
Log-transformed and normalized data before and after BEC are shown (see Methods section for details on data preprocessing). For each dataset, sample projections to the first two PCs are shown. The percentage of explained variance is shown, demonstrating the drop in variance explained by batch after BEC.
Panels: **a.** *E. coli* proteomics dataset. **b.** Ovarian cancer microarray dataset. **c.** ccRCC proteomics data. **d.** Quartet multi-omics dataset.

## Federated BEC improves the outcomes of FL

After confirming that fedRBE produces results consistent with centralized BEC using limma's removeBatchEffect() across datasets and varying levels of class imbalance, we next evaluated how batch effects affected the performance of supervised and unsupervised FL models. We applied federated random forest (RF; see Methods) and federated k-means[58] to discriminate between known biological groups in each of the four datasets, and compared model performances before and after BEC. For supervised learning, we considered two train-test splitting scenarios: (i) a within-client split, in which the data from each client were divided into training and test sets at a 70:30 ratio, and (ii) a leave-one-center-out scenario, in which the model was trained on data from all but one client and evaluated on the held-out client.

In many experiments, BEC of decentralized data using fedRBE substantially improved model performance relative to FL without BEC (**Figure 4**). This improvement was particularly pronounced for unsupervised learning, where BEC was critical for recovering biologically meaningful structure (**Figure 4a**). Unlike RF, standard k-means does not have a mechanism to account for batch-related variation. Therefore, when batch effects are not removed, clustering could be driven by technical differences between clients rather than by the biological groups of interest. In contrast to standard k-means applied to pooled data, its federated version by design leverages information that samples originate from distinct clients (see Methods). This feature can improve algorithmic performance and convergence in some cases relative to standard k-means, but it does not fully compensate for the absence of prior BEC.

In the supervised classification task, BEC substantially improved classification performance in the leave-one-center-out scenario in three of the four datasets (**Figure 4b**). For example, in the *E. coli* classification task under the leave-one-out scenario, applying BEC increased the median MCC from 0, corresponding to random prediction, to 1, indicating perfect classification. In addition to improving the MCC on average, applying BEC before federated RF training reduced the variability of the results and appeared to facilitate more stable model convergence. The only exception was the Quartet dataset, in which all models achieved perfect performance irrespective of BEC. This dataset exhibited the lowest batch-specific variance, suggesting that it may contain discriminative features that are relatively insensitive to batch effects.

In the within-client 70:30 test split scenario, the same tasks were solved nearly perfectly both with and without prior BEC. This marked difference between two validation scenarios is likely explained by the fact that, in the leave-one-center-out scenario, the model has no direct access during training to the batch-specific characteristics of the held-out client's data, making it more vulnerable to batch effects.

a) k-means clustering

b) RF leave-one-cohort-out

c) RF 70:30 train-test split

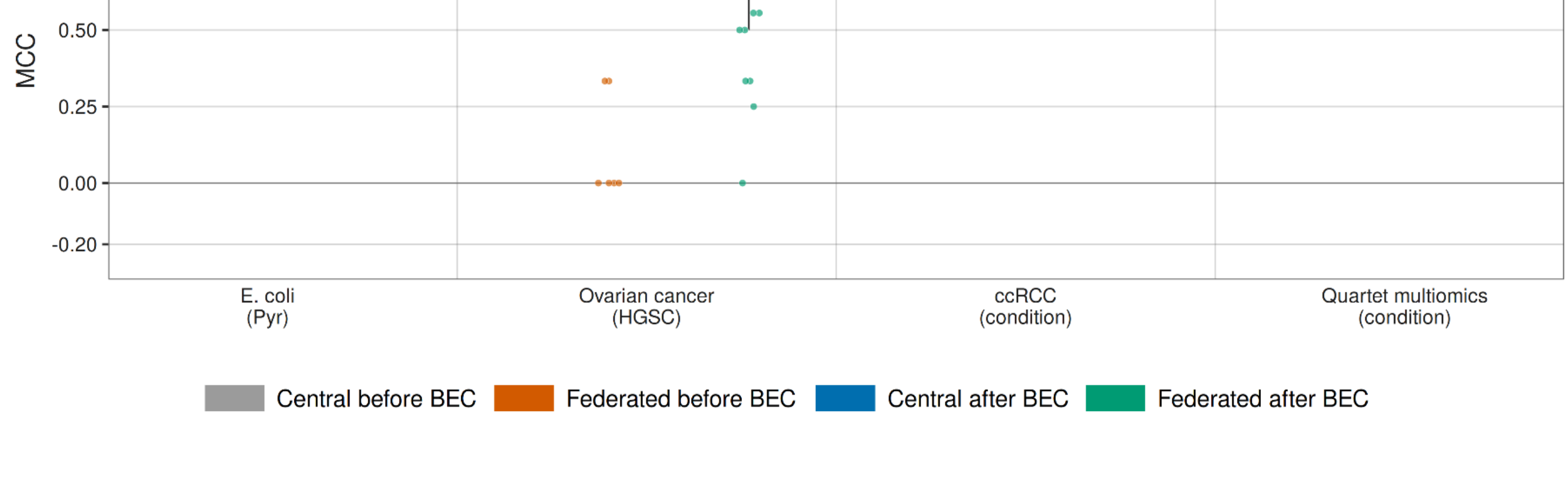

**Figure 4**. Performance of centralized and federated models in unsupervised **(a)** and supervised **(b,c)** learning tasks before and after batch effect correction (BEC). For each dataset/task, four scenarios were compared: centralized models on pooled uncorrected data (grey), federated models on decentralized uncorrected data (orange), centralized models on pooled data corrected with limma's *removeBatchEffect*() (blue), and federated models on decentralized BEC-corrected data (green).
**a.** Adjusted Rand Index (ARI) for clusters identified by standard and federated k-means. For each dataset, clustering was repeated 10 times with different random seeds. **b,c.** Matthews correlation coefficient (MCC) for classes predicted by centralized and federated RF models under leave-one-out (**b**) and 70:30 train-test split (**c**) scenarios. Each split scenario was repeated 10 times with different random seeds, resulting in $10 \times c$ total trained models for each dataset, where $c$ is the number of centers. Each datapoint in the boxplot represents the performance on a client's test data. Under the 70:30 train–test split (c) this means that for one RF training $c$ MCC values are recorded. The 70:30 split seed affects both data splitting and RF training.

## Discussion

Batch effects in omics data typically result in substantial systematic shifts of feature distributions across different centers, with magnitudes that can substantially exceed those of the biological signals of interest. In this study, we directly evaluated the impact of batch effects on both supervised and unsupervised FL, using four multi-center datasets spanning different omics types.

Since unsupervised learning methods by definition do not use any sample labels, including indicators of sample source site, they are particularly vulnerable to batch effects. When batch variance largely exceeds biological variance, $k$-means with Euclidean distance is expected to partition samples primarily by batch. Consistent with this expectation, in the absence of BEC, federated $k$-means was unable to identify the target classes. After BEC, clustering performance improved substantially, with median ARI values increasing to 0.7-1.0 across the datasets.

Supervised learning methods can utilize the information that data originate from different centers and are therefore capable of learning batch effects, although this requires increasing model complexity. In the case of a federated random forest, the model may learn center-specific features and thresholds that optimize discrimination of the target classes within individual centers. If a model lacks a mechanism that forces it to select the same features across all centers, this can potentially lead to overfitting.

Furthermore, the impact of batch effects on supervised model performance is also dependent on the specific training scenario. When one or more centers are entirely absent from the training set, the model cannot directly learn batch effects specific to these unseen centers. Under such conditions, successful prediction is possible only if the model captures universal discriminative patterns that generalize across centers, or if the batch effect is relatively weak.

Beyond gains in average performance, addition of BEC also tended to stabilize model performance, with the greatest variability observed in unsupervised and leave-one-center-out supervised learning scenarios. These results suggest that harmonizing data across centers before FL can improve its performance and robustness, thereby motivating practical privacy-preserving implementations of federated BEC.

To facilitate the adoption of federated BEC by the research community, we created fedRBE, a privacy-preserving version of the popular batch effect removal method removeBatchEffect() from the *limma* R package[19]. The privacy of each participant's data is protected by sharing only aggregated intermediate results, further enhanced with SMPC, and by detecting edge cases with insufficient sample sizes. We validated fedRBE using both real and simulated data and compared its results with the results of *limma* applied to the same data collected centrally. In all tests, we observed negligible numerical differences arising from floating-point arithmetic differences and their propagation across the multi-step computation between central and federated implementations.

To our knowledge, fedRBE is the first fully automated BEC tool for bulk omics data that supports inputs with missing values and non-identical feature sets across clients, models covariates, and provides a user-friendly browser-based GUI, enabling decentralized BEC for users with diverse computational backgrounds. FedRBE is available for users as a FeatureCloud[47] application (https://featurecloud.ai/app/fedrbe), and can be used independently or be incorporated into various decentralized workflows in combination with other FeatureCloud applications, enabling FL workflows that were previously infeasible due to batch effects.

While fedRBE enables privacy-aware federated BEC and facilitates its integration into decentralized data analysis workflows, several limitations of this study should be considered. First, because fedRBE is based on *limma*'s removeBatchEffect() method, it inherits its application scope and limitations. Some data types, such as count-based RNA-seq data, may require a different approach to BEC[29]. In addition, *limma* relies on a linear modeling framework, whereas batch effects in real data may not always be adequately described by a linear transformation[59]. Therefore, future work could be focused on developing privacy-aware BEC tools tailored to specific data distributions and capable of correcting non-linear batch effects.

Second, we evaluated the effect of BEC using two representative FL algorithms, federated $k$-means and federated random forest, whereas more specialized approaches, including ensemble weighting[60], may demonstrate greater robustness to batch effects and require further investigation.

Finally, BEC should be applied cautiously in unsupervised settings, particularly when unknown biological groups are unevenly distributed across centers, because unaccounted-for confounding between batch and biological structure may result in overcorrection erasing genuine biological variation.

# Methods

## The fedRBE algorithm

The federated BEC involves $c$ participants, with one of them also designated as the coordinator. The participants' clients hold local datasets, with each client $i \in \{1, ..., c\}$ possessing a local design matrix $X_i \in \mathbb{R}^{n_i \times h}$, where $h$ columns represent $h$ covariates, and a local data matrix $Y_i \in \mathbb{R}^{m_i \times n_i}$, whith $m_i$ features (e.g., genes) and $n_i$ samples. While all clients need to hold the same covariates, fedRBE supports different sets of features for each client. To ensure the functionality of the SMPC protocol used[46], features must be available in at least three clients (**Figure 2**). Additionally, fedRBE requires covariates to be linearly independent from each other and the batches.

FedRBE performs federated computations and produces results mathematically equivalent to the results of *limma*'s removeBatchEffect()[19] applied to the pooled dataset.

At the start of the workflow, clients are assigned random identifiers, establishing a randomized client sequence. Batches are first arranged according to this client sequence, with batches belonging to the same client then ordered alphanumerically. If desired, users may specify a fixed client sequence for a deterministic batch order. The last batch in this final sequence is designated as the reference batch. Alternatively, one client can specify one of their batches to be used as the reference batch. The batch information, together with an intercept column, is added to the client's $X_i$, forming $X_i \in \mathbb{R}^{n_i \times (1+h+k-1)}$, where the first column is an intercept column, the following $h$ columns are the covariates, and the last $k - 1$ columns are the columns representing sample membership in $k$ batches, with the reference batch not represented as a column but as a value of minus one in the other columns.

Furthermore, all clients share the names of the covariates and features per batch with the coordinator. To conceal covariates and features unknown to the coordinator, all covariates and features are hashed. Then, the intersection of all given covariates and the union of all features are selected by the coordinator and shared with all clients. Features only available in two or fewer clients are removed from this union.

FedRBE implements federated linear regression as proposed by Karr et al.[61]. To handle missing values in $y_j$, the entries of patients with missing values are removed from both $X_{ij}$ and $y_{ij}$ before any further calculations, as they cannot contribute to the final model. Therefore, to avoid rank deficiency problems emerging due to missing data in the entire batch of any $y_{ij}$, a newly developed masking approach is applied to remove corresponding columns from $X_i$ and $y_i$ (Supplementary Methods and Supplementary Figure S6). *Limma's removeBatchEffect()* method handles missing values in the same manner.

To obtain the global $X_j^T X_j$ and $X_j^T y_j$ matrices, each client $i$ calculates their local $X_{ij}^T X_{ij}$ and $X_{ij}^T y_{ij}$ for each feature $j$ . The coordinator then adds up the received matrices of all $c$ clients:

$$X_j^T X_j = \sum_{i=1}^{c} X_{ij}^T X_{ij} \quad ,$$

$$X_j^T y_j = \sum_{i=1}^{c} X_{ij}^T y_{ij} \quad .$$

These sums are calculated securely using SMPC (see below), so the coordinator has access only to the final $X_j^T X_j$ and $X_j^T y_j$ without accessing any local $X_{ij}^T X_{ij}$ and $X_{ij}^T y_{ij}$. Local $X_i$ and $Y_i$ are never transmitted.

The global model coefficients $\widehat{\beta}_j$ are then calculated as $\hat{\beta}_j = (X_j^T X_j)^{-1} X_j^T y_j$, ignoring batches according to the mask $D_j$, and shared with the clients, who further correct their local data:

$$y'_{ij} = y_{ij} - X_{ij}\hat{\beta}_j.$$

## Enhancing privacy with additive secret sharing

To protect local $X_{ij}^T X_{ij}$ and $X_{ij}^T y_{ij}$, the additive secret sharing scheme implemented in FeatureCloud is used[39,46]. In the additive secret sharing technique, any set of numeric data of the same dimensions can be aggregated without knowing the individual set members. For any numerical data piece, e.g., for the matrix $X_{ij}^T X_{ij}$, $c - 1$ data pieces of the same dimensionality are generated by drawing random integers from $\left[- 2^{43}, 2^{43}\right]$. Then, each client creates the last data piece by subtracting the sum of the random data pieces from the original data. This ensures that the $c$ data pieces of the same client sum up to the original data. All but one of the data pieces are further distributed across all other $c - 1$ clients, resulting in each client holding $c$ pieces in total. Finally, each client then aggregates the $c$ pieces and sends the aggregate to the coordinator, which then combines the aggregated data from all clients.

## Impact of batch effects on FL performance

To evaluate the impact of batch effects on downstream FL, we assessed two widely used machine learning approaches representing unsupervised and supervised settings: k-means clustering and random forest classification. For both methods, we compared the ability of the algorithms to recover biological classes defined in the sample metadata before and after BEC.

Federated implementations of k-means clustering and random forest classification were applied to the same datasets before and after batch effect correction using FedRBE. The adjusted Rand index (ARI) and the Matthews correlation coefficient (MCC) calculated for predicted and true sample labels were used to quantify clustering and classification performances, respectively. To assess the robustness of the results, each algorithm was run 10 times with the same input data using different random seeds.

In addition, we compared the performance of the federated algorithms with that of conventional centralized implementations applied to the corresponding pooled datasets. For the centralized analysis, BEC was performed using the *removeBatchEffect*() function from the limma R package.

### Random forest

Random forest classification was performed using the implementation available in the scikit-learn Python package v1.8.0. The maximal tree depth was limited to 10, and 75% of samples were used to train each individual estimator. All remaining parameters were kept at their default values, including the maximum number of estimators, which was set to 100.

Federated random forest classification was performed by training local RF models independently on each client's data and subsequently aggregating them into a single global ensemble. To keep the size of the global RF approximately equal to 100 trees, each client trained $\lceil 100/c \rceil$ trees, where c denotes the number of participating centers.

For supervised learning, model performance was evaluated under two distinct training scenarios. In the leave-one-center-out scenario, models were trained using data from c − 1 centers, while data from the remaining center was reserved for evaluation. In the 70:30 train-test split scenario, each center randomly reserved 30% of its local samples for evaluation, while the remaining 70% were used for model training. Each train-test split was controlled by a random seed, and the composition of the training and evaluation subsets differed across repeated runs.

### k-means clustering

Before clustering, data were standardized across samples with sklearn.preprocessing.StandardScaler from scikit-learn v1.3.2 Python package. For each dataset, features containing missing values were excluded, as required by k-means implementations.

The number of clusters was set to the number of biological target classes for the corresponding dataset. The centralized workflow used the Python sklearn.cluster.KMeans implementation with n_init = 1 by default; the centralized k-means was run in Python 3.11.14 with scikit-learn 1.8.0.

Federated k-means was run with the FeatureCloud Federated K-Means workflow[58] (https://featurecloud.ai/app/fc-federated-kmeans). The implemented algorithm is a single-shot approach which was adapted from[62]. In this workflow, the data centers initialize local centroids using sklearn.KMeans with a fixed k which is run until convergence and share the centroid summaries with the coordinator. The coordinator aggregates the centroids by running sklearn.KMeans on the centroids with the same fixed k and shares the aggregated global centroids with the data holders. The clients assign the local samples to the nearest centroid resulting in a clustering where every datapoint is assigned to the global centroid without having shared the data points.

## Data preprocessing

All real-world evaluations used public multi-center datasets with biological target labels and center- or study-specific batch labels. For each dataset, the same biological covariates and batch definitions were used for centralized BEC and for fedRBE.

### *E. coli* proteomics dataset

A multi-center mass spectrometry-based proteomics dataset consisting of 118 *Escherichia coli* samples from five laboratories was obtained from the PRIDE Archive, PXD053812[28].

The data were acquired using data-independent acquisition label-free quantitative proteomics and were preprocessed using the MaxLFQ algorithm[63] implemented in DIA-NN[64] (v1.8.1). We filtered the dataset to keep only 118 samples from the two conditions, grown on pyruvate and glucose media, and performed a log2 transformation of normalized intensities. The laboratory of origin was used as the batch factor and growth condition as the biological target.

### Ovarian cancer microarray dataset

The aggregated microarray gene expression dataset consisted of six publicly accessible ovarian cancer datasets from the Gene Expression Omnibus, accession numbers GSE6008[48], GSE26712[49], GSE40595[50], GSE69428[51], GSE38666[52], and GSE14407[53].

From each study, control and high-grade serous carcinoma (HGSC) samples were selected and preprocessed independently using the robust multi-array average (rma) method from the *affy* R package[65] (v. 1.78.0). The data were subsequently annotated and aggregated to gene-level expressions using the “maxRowVariance” method from the collapseRows() function in the *WGCNA* R package[66] (v. 1.71). The source study was used as the batch factor and disease status as the biological target.

### ccRCC proteomics dataset

The ccRCC proteomics dataset was assembled from three public studies, PDC000127[54], PXD030344[55], and PXD042844[56]. For each study, the publicly available filtered protein-intensity matrix was used as the site-level input without additional preprocessing. The source study was used as the batch factor and disease status as the target variable.

### Quartet multi-omics dataset.

To simulate three clients with multi-center multi-omics data, we used a subset of transcriptomic, proteomic, and metabolomic profiles from the Quartet dataset[57]. This dataset comprises molecular profiles generated from reference materials derived from lymphoblastoid cell lines derived from four members of a single family: monozygotic twin daughters D5 and D6, and their parents F7 and M8. It provides repeated measurements of four reference samples across multiple omics layers, platforms, laboratories and experimental batches.

The preprocessed Quartet data were downloaded from Figshare[67] and used as provided, without additional modification. Each selected batch contained three replicates for each of the four Quartet reference samples, resulting in 12 profiles per modality per batch. As in the other tests, batch labels were defined according to the experimental batch annotations and individual identity labels (D5, D6, F7, or M8) were used as biological condition labels.

The first two clients received one experimental batch each, with 12 transcriptomic, 12 proteomic, and 12 metabolomic profiles per client, measured by laboratories L01 and L02. The third simulated client comprised two experimental batches. For transcriptomic and proteomic data, both batches were obtained from laboratory L05. For metabolomic data, one batch from laboratory L05 and one batch from laboratory L04 were used (**Supplementary Table S1**).

Central and federated BEC were run separately within each modality. For the joint multi-omics RF and k-means analyses, the corrected transcriptomics, proteomics, and metabolomics matrices were stacked by feature into matched pseudo-samples defined by client, donor, replicate, and within-cell library index.

## Simulated datasets

Additionally, to investigate the effect of data imbalance on fedRBE results, we adapted and extended the simulated data generation approach described in Burankova et al., 2025[28]. We used the same number of samples, batch distribution, and scenarios (balanced, mild imbalanced, and strong imbalanced data; see **Supplementary Table S3**).

We simulated datasets with three levels of imbalance: a balanced dataset with equal sample counts across centers and conditions (no imbalance), a mildly imbalanced dataset with moderate variations in sample counts between conditions and centers, and a strongly imbalanced dataset. Following the general framework described by Wang et al. and implemented in the *RobNorm* R package[68], we modified the parameters of the simulation. Background data were simulated with means $\mu_j$ drawn from a standard normal distribution and variances $\sigma_j^2$ from an inverse gamma distribution $InvGamma(2, 3)$. We introduced a greater number of differentially expressed features. The mean shift (Δμ) for the differentially expressed features was adjusted to 1.25 for 2,250 features and to 1.25 for 150 confounder features.

The parameters for batch effects modeling using the ComBat model[20] were also adjusted: additive effects were drawn from normal distributions with means $\{0, 0.7, -1.5\}$ and standard deviations $\{0.5,\ 1,\ 1.5\}$, while multiplicative effects were modeled using gamma distributions with shape parameters $\{3,\ 1,\ 5\}$ and scale parameters $\{2, 1, 0.5\}$, corresponding to each of the three batches, respectively. Between batches, additive and multiplicative effects differed by at least 20%. The additive and multiplicative batch parameters were kept consistent across all features within a batch.

## Evaluation of BEC results

For all described datasets, variance partitioning was used to quantify the impact of batch effect correction on data[69]. The analysis was performed using the fitExtractVarPartModel() function from the *variancePartition* R package (v. 1.30.2), which evaluates the contribution of each aspect of the study design to the variance of each feature (gene, protein group, etc.). It fits a linear model to estimate the proportion of variance in the data attributable to different sources. Each entry in the analysis results is a variance fraction extracted from a regression model fitted on a single feature. It can be interpreted as the variance explained by each variable after correcting for all other variables.

Principal component analysis (PCA) plots were created in R version 4.3.2 (2023-10-31), using *ggplot2* v.3.5.1[70] R package. The plots were aggregated in BioRender.

# Acknowledgements

Figures 1 and 2 were created by BioRender.com.

# Funding

Funded by the European Union under contract no. 101136305. The Hungarian partner is funded by the Hungarian National Research, Development, and Innovation Fund. Views and opinions expressed are however those of the author(s) only and do not necessarily reflect those of the European Union or the Hungarian National Research, Development and Innovation Fund. Neither the European Union nor the Hungarian National Research, Development and Innovation Fund can be held responsible for them. The Microb-AI-ome project has received funding from the European Union's Horizon research and innovation programme under the Grant Agreement nº 101079777. The information contained in this paper are however those of the authors only and do not necessarily reflect those of the European Union. This work was also developed as part of the FeMAI project and is funded by the German Federal Ministry of Research, Technology and Space (BMFTR) under grant number 16IS21079. This work was also funded by the German Federal Ministry of Research, Technology and Space (BMFTR) and the Free and Hanseatic City of Hamburg under the Excellence Strategy of the Federal Government and the Länder. Last, this work was supported by the German Federal Ministry of Research, Technology and Space (BMFTR) within the framework of "CLINSPECT-M-2" (grant 16LW0243K). This research was supported by CVDLINK project. This project was awarded funding by the European Union's Horizon Research and Innovation programme under grant agreement N°101137278. Views and opinions expressed are however those of the author(s) only and do not necessarily reflect those of the European Union or the European Health and Digital Executive Agency. Neither the European Union nor the granting authority can be held responsible for them.

# Conflicts of Interest

The authors declare that they have no competing interests.

# Data Availability

All data used for evaluation are publicly available. The raw mass spectrometry data for the MS-based proteomics *E. coli* dataset [28] are available on the PRIDE repository with the dataset identifiers PXD053812[71]. The raw data for the microarray-based transcriptomics ovarian cancer dataset are available via Gene Expression Omnibus[72,73], accession numbers GSE6008[48], GSE26712[49], GSE40595[50], GSE69428[51], GSE38666[52], GSE14407[53]. The ccRCC proteomics data are available under PDC000127[54], PXD030344[55], and PXD042844[56]. The Quartet multi-omics data are available from Figshare[67].

# Code Availability

FedRBE is freely accessible through its web interface[47]. Registration is required in a multi-center analysis of sensitive human data and is necessary to manage invitation tokens, prevent unauthorized access, and legal and institutional compliance, including GDPR obligations, access-controlled collaboration.

The FeatureCloud App, as well as scripts to reproduce the results and analyses presented in this article, can be found on GitHub (https://github.com/Freddsle/fedRBE) and Zenodo[74]. The repository also contains scripts that locally simulate federated learning, for which the web interface may be used for managing the simulated BEC but no registration is required.

# Authors' contributions

Y.B., O.Z. — fedRBE algorithm development. Y.B., J.K., J.J.G.L — fedRBE app implementation and testing. Y.B., J.K., A.T., N.P. — real-world datasets preprocessing and fedRBE evaluation on them. Y.B. – unsupervised learning evaluation. J.K. – supervised learning evaluation. Y.B., J.K. — simulated datasets preprocessing and evaluation on them. Y.B., O.Z. — designed the study. J.B., O.Z. — supervised the work. Y.B., J.K., J.J.G.L., O.Z. — drafted the manuscript and designed the figures (with input from all authors).

All authors have read and approved the final version of the manuscript.

# Abbreviations

AI: artificial intelligence
ARI: adjusted Rand index
BEC: batch effect correction
CCPA: California Consumer Privacy Act
ccRCC: clear cell renal cell carcinoma
CSV: comma-separated values
FL: federated learning
GDPR: General Data Protection Regulation
GUI: graphical user interface
HGSC: high-grade serous carcinoma
HIPAA: Health Insurance Portability and Accountability Act
MCC: Matthews correlation coefficient
MS: mass spectrometry
NA: missing value indicator (“Not available”)
PCA: principal component analysis
RF: random forest
SMPC: secure multi-party computation

# Supplementary information

## Batch effects can impair federated learning in multi-center omics studies

Yuliya Burankova[1,2,*,†], Julian Klemm[1,†], Jens J. G. Lohmann[1], Anne Hartebrodt[3,4], Ahmad Taheri[1], Niklas Probul[1], Jan Baumbach[1,3,‡], Olga Zolotareva[1,5,6,*,‡]

[1] *Institute for Computational Systems Biomedicine, University of Hamburg, 22761 Hamburg, Germany*
[2] *Chair of Proteomics and Bioanalytics, TUM School of Life Sciences, Technical University of Munich, 85354 Freising, Germany*
[3] *Department of Mathematics and Computer Science, University of Southern Denmark, 5230 Odense, Denmark*
[4] *Biomedical Network Science Lab, Department Artificial Intelligence in Biomedical Engineering, Friedrich-Alexander-Universität Erlangen-Nürnberg, 91052 Erlangen, Germany*
[5] *Data Science in Systems Biology, TUM School of Life Sciences, Technical University of Munich, 85354 Freising, Germany*
[6] Institute of Clinical Molecular Biology (IKMB), Kiel University and University Medical Center Schleswig-Holstein, Rosalind-Franklin-Str. 12, 24105 Kiel, Germany

*To whom correspondence should be addressed:
Yuliya Burankova, Email: yuliya.burankova@tum.de
Julian Klemm, Email: julian.klemm@uni-hamburg.de
Correspondence may also be addressed to Olga Zolotareva. Email: o.zolotareva@ikmb.uni-kiel.de
†The first two authors should be regarded as Joint First Authors.
‡The last two authors should be regarded as Joint Last Authors.

**Content:**

# Supplementary Methods

## Masking technique

To avoid rank deficiency problems emerging due to an entire batch missing data, fedRBE uses a masking approach, eliminating coefficients in $\beta_j$ of federated linear regression [1].

To construct the global mask, we first check the presence of each feature to determine its availability in each batch $k$. We construct a matrix $D \in \mathbb{R}^{m \times (1+h+k-1)}$ where each row is a mask for the feature $j$. The mask $D_j$ indicates which batches should be included for feature $j$ in the model (**Supplementary Figure S6**). This replicates the computation process in *removeBatchEffect()* and ensures that for each feature $j$, any potential linear dependencies due to missing data are mitigated.

# Supplementary Tables

**Table S1.** Center-level composition of the real-world datasets used for fedRBE evaluation.

<table>
<tr><th>Dataset</th><th>Center / study / client</th><th>Modality</th><th>Features</th><th>Samples</th><th>Target counts</th></tr>
<tr><td rowspan="5"><b>E. coli</b></td><td>lab_A</td><td rowspan="5">Proteomics</td><td>2549</td><td>24</td><td>glucose 12; pyruvate 12</td></tr>
<tr><td>lab_B</td><td>2846</td><td>23</td><td>glucose 11; pyruvate 12</td></tr>
<tr><td>lab_C</td><td>2820</td><td>23</td><td>glucose 12; pyruvate 11</td></tr>
<tr><td>lab_D</td><td>2813</td><td>24</td><td>glucose 12; pyruvate 12</td></tr>
<tr><td>lab_E</td><td>2401</td><td>24</td><td>glucose 12; pyruvate 12</td></tr>
<tr><td rowspan="6"><b>Ovarian cancer</b></td><td>GSE6008</td><td rowspan="6">Transcriptomics</td><td rowspan="2">21128</td><td>27</td><td>controls 4;<br>HGSC tumors 23</td></tr>
<tr><td>GSE26712</td><td>195</td><td>controls 10;<br>HGSC tumors 185</td></tr>
<tr><td>GSE40595</td><td rowspan="4">51276</td><td>37</td><td>controls 6;<br>HGSC tumors 31</td></tr>
<tr><td>GSE69428</td><td>19</td><td>controls 9;<br>HGSC tumors 10</td></tr>
<tr><td>GSE38666</td><td>30</td><td>controls 12;<br>HGSC tumors 18</td></tr>
<tr><td>GSE14407</td><td>24</td><td>controls 12;<br>HGSC tumors 12</td></tr>
<tr><td rowspan="3"><b>ccRCC</b></td><td>PDC000127</td><td rowspan="3">Proteomics</td><td>9964</td><td>194</td><td>controls 84; tumors 110</td></tr>
<tr><td>PXD030344</td><td>12548</td><td>464</td><td>controls 232; tumors 232</td></tr>
<tr><td>PXD042844</td><td>7655</td><td>229</td><td>controls 114; tumors 115</td></tr>
<tr><td rowspan="9"><b>Quartet multi-omics</b></td><td>client_01_L01</td><td rowspan="3">Transcriptomics</td><td rowspan="3">26907</td><td>12</td><td rowspan="2">D5 3; D6 3; F7 3; M8 3</td></tr>
<tr><td>client_02_L02</td><td>12</td></tr>
<tr><td>client_03_L05_L04</td><td>24</td><td>D5 6; D6 6; F7 6; M8 6</td></tr>
<tr><td>client_01_L01</td><td rowspan="3">Proteomics</td><td rowspan="3">2539</td><td>12</td><td rowspan="2">D5 3; D6 3; F7 3; M8 3</td></tr>
<tr><td>client_02_L02</td><td>12</td></tr>
<tr><td>client_03_L05_L04</td><td>24</td><td>D5 6; D6 6; F7 6; M8 6</td></tr>
<tr><td>client_01_L01</td><td rowspan="3">Metabolomics</td><td rowspan="3">71</td><td>12</td><td rowspan="2">D5 3; D6 3; F7 3; M8 3</td></tr>
<tr><td>client_02_L02</td><td>12</td></tr>
<tr><td>client_03_L05_L04</td><td>24</td><td>D5 6; D6 6; F7 6; M8 6</td></tr>
</table>

**Table S2.** Absolute differences (errors) between the results of batch effect correction using *limma* removeBatchEffect() applied on aggregated datasets and the results of fedRBE applied to the same decentralized data.

| Dataset | SMPC | Minimal absolute error | Mean absolute error | Maximal absolute error |
|---|---|---|---|---|
| *E. coli* proteomics dataset | no | 0 | 3.24E-14 | 1.92E-13 |
| | yes | 0 | 3.39-14 | 2.49E-13 |
| Microarray ovarian cancer transcriptomics dataset | no | 0 | 3.89E-15 | 7.99E-14 |
| | yes | 0 | 1.50E-14 | 1.58E-13 |
| ccRCC proteomics dataset | no | 0 | 1.60E-15 | 5.15E-14 |
| | yes | 0 | 2.79E-15 | 1.28E-13 |
| Quartet multi-omics transcriptomics | no | 0 | 1.92E-15 | 3.73E-14 |
| | yes | 0 | 2.80E-14 | 1.47E-13 |
| Quartet multi-omics proteomics | no | 0 | 2.12E-15 | 3.73E-14 |
| | yes | 0 | 2.96E-14 | 1.50E-13 |
| Quartet multi-omics metabolomics | no | 0 | 9.44E-15 | 4.09E-14 |
| | yes | 0 | 2.15E-14 | 1.01E-13 |

**Table S3.** Characteristics of the simulated datasets used to evaluate the effect of data imbalance (adapted from Burankova et al., 2025 [2]). Number of samples in each cohort in each condition, for the confounder column — proportion of samples among condition B samples.

| | Cohorts | | | Condition A | | | Condition B | | | in B — frequency of samples with the confounder | | |
|---|---|---|---|---|---|---|---|---|---|---|---|---|
| | B 1 | B 2 | B 3 | B 1 | B 2 | B 3 | B 1 | B 2 | B 3 | B 1 | B 2 | B 3 |
| **Balanced** | 200 | 200 | 200 | 100 | 100 | 100 | 100 | 100 | 100 | 0.6 | 0.6 | 0.6 |
| **Mild imbalanced** | 90 | 140 | 370 | 36 | 91 | 185 | 54 | 49 | 185 | 0.4 | 0.5 | 0.66 |
| **Strong imbalanced** | 40 | 80 | 480 | 32 | 28 | 288 | 8 | 52 | 192 | 0.2 | 0.5 | 0.7 |

**Table S4.** Absolute differences (errors) between the results of batch effect correction using *limma* removeBatchEffect() applied on aggregated datasets and the results of fedRBE applied to the same decentralized data. The results are shown for 30 simulation runs of three scenarios of simulated data together.

| Dataset | Mean error | Maximal error |
|---:|:---:|:---:|
| Balanced datasets | 2.31e-15 | 4.97e-13 |
| Mildly imbalanced datasets | 2.26e-15 | 5.68e-14 |
| Strongly imbalanced datasets | 2.04e-15 | 2.70e-13 |

# Supplementary Figures

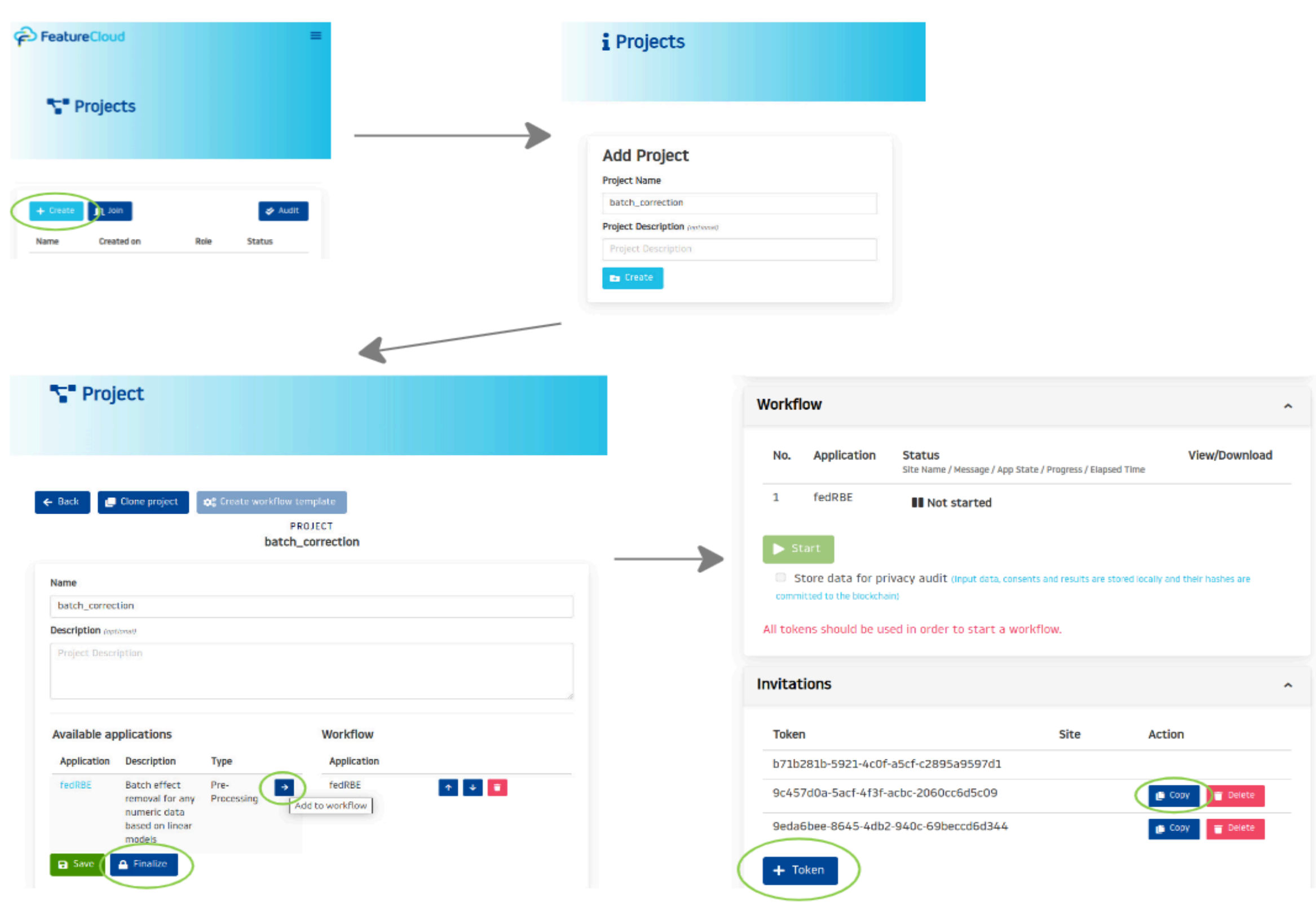

**Figure S1.** The overview of the coordinator steps for creating the fedRBE project and inviting participants. The coordinating user creates a project, adds the fedRBE app into it, and finalizes the workflow. After that, the coordinating user creates invitation tokens for the participants and distributes them.

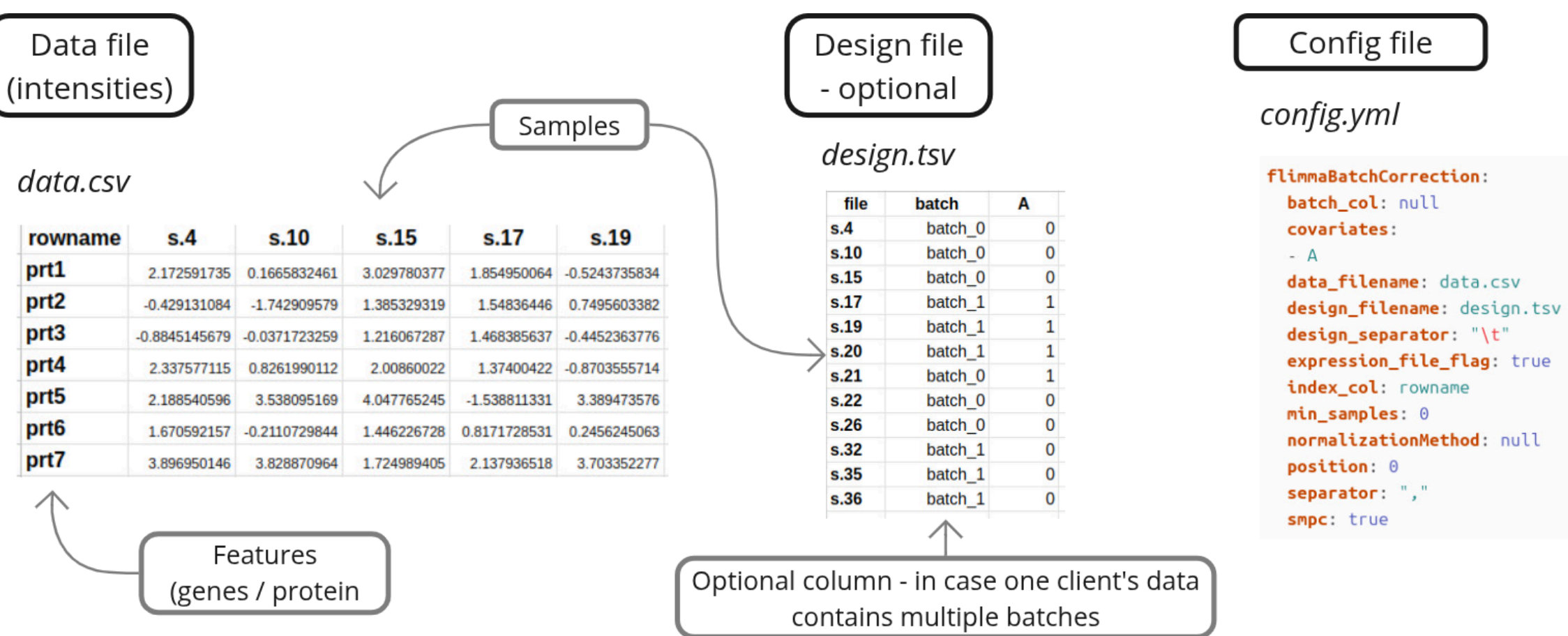

| rowname | s.4 | s.10 | s.15 | s.17 | s.19 |
|---|---|---|---|---|---|
| prt1 | 2.172591735 | 0.1665832461 | 3.029780377 | 1.854950064 | -0.5243735834 |
| prt2 | -0.429131084 | -1.742909579 | 1.385329319 | 1.54836446 | 0.7495603382 |
| prt3 | -0.8845145679 | -0.0371723259 | 1.216067287 | 1.468385637 | -0.4452363776 |
| prt4 | 2.337577115 | 0.8261990112 | 2.00860022 | 1.37400422 | -0.8703555714 |
| prt5 | 2.188540596 | 3.538095169 | 4.047765245 | -1.538811331 | 3.389473576 |
| prt6 | 1.670592157 | -0.2110729844 | 1.446226728 | 0.8171728531 | 0.2456245063 |
| prt7 | 3.896950146 | 3.828870964 | 1.724989405 | 2.137936518 | 3.703352277 |

| file | batch | A |
|---|---|---|
| s.4 | batch_0 | 0 |
| s.10 | batch_0 | 0 |
| s.15 | batch_0 | 0 |
| s.17 | batch_1 | 1 |
| s.19 | batch_1 | 1 |
| s.20 | batch_1 | 1 |
| s.21 | batch_0 | 1 |
| s.22 | batch_0 | 0 |
| s.26 | batch_0 | 0 |
| s.32 | batch_1 | 0 |
| s.35 | batch_1 | 0 |
| s.36 | batch_1 | 0 |

```yaml
flimmaBatchCorrection:
  batch_col: null
  covariates:
  - A
  data_filename: data.csv
  design_filename: design.tsv
  design_separator: "\t"
  expression_file_flag: true
  index_col: rowname
  min_samples: 0
  normalizationMethod: null
  position: 0
  separator: ","
  smpc: true
```

**Figure S2.** Example structure of client input files. The parameters in *config.yml* can be adjusted as necessary, for example, by updating the data_filename to match the actual file name. In cases where there are multiple batches per client, or any covariates must be taken into account, the participants should include a *design.csv* file containing the relevant information and modify their *config.yml* file accordingly (under the parameter batch_col for multiple batches).

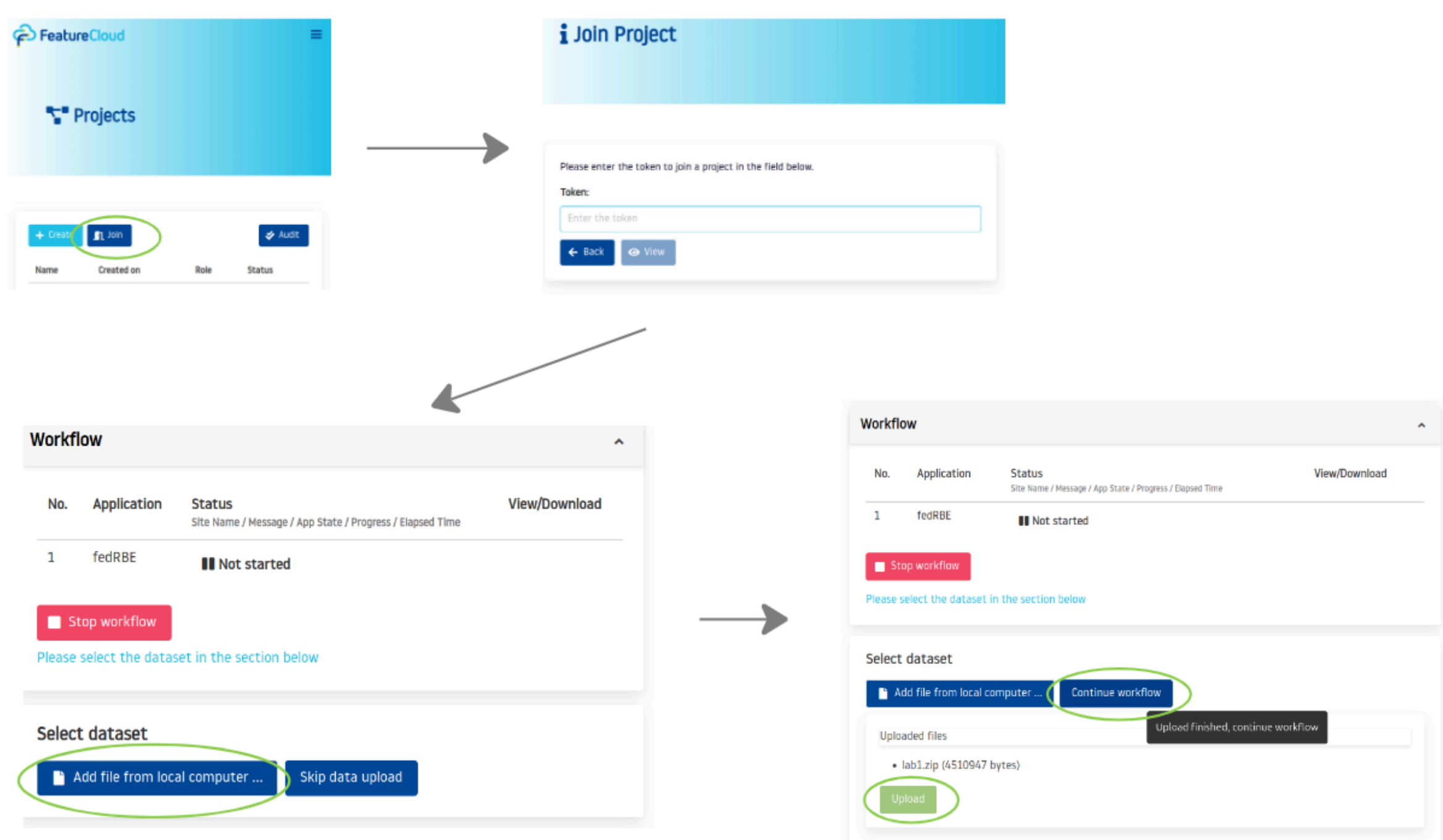

**Figure S3.** The overview of client steps for joining the analysis and providing data.

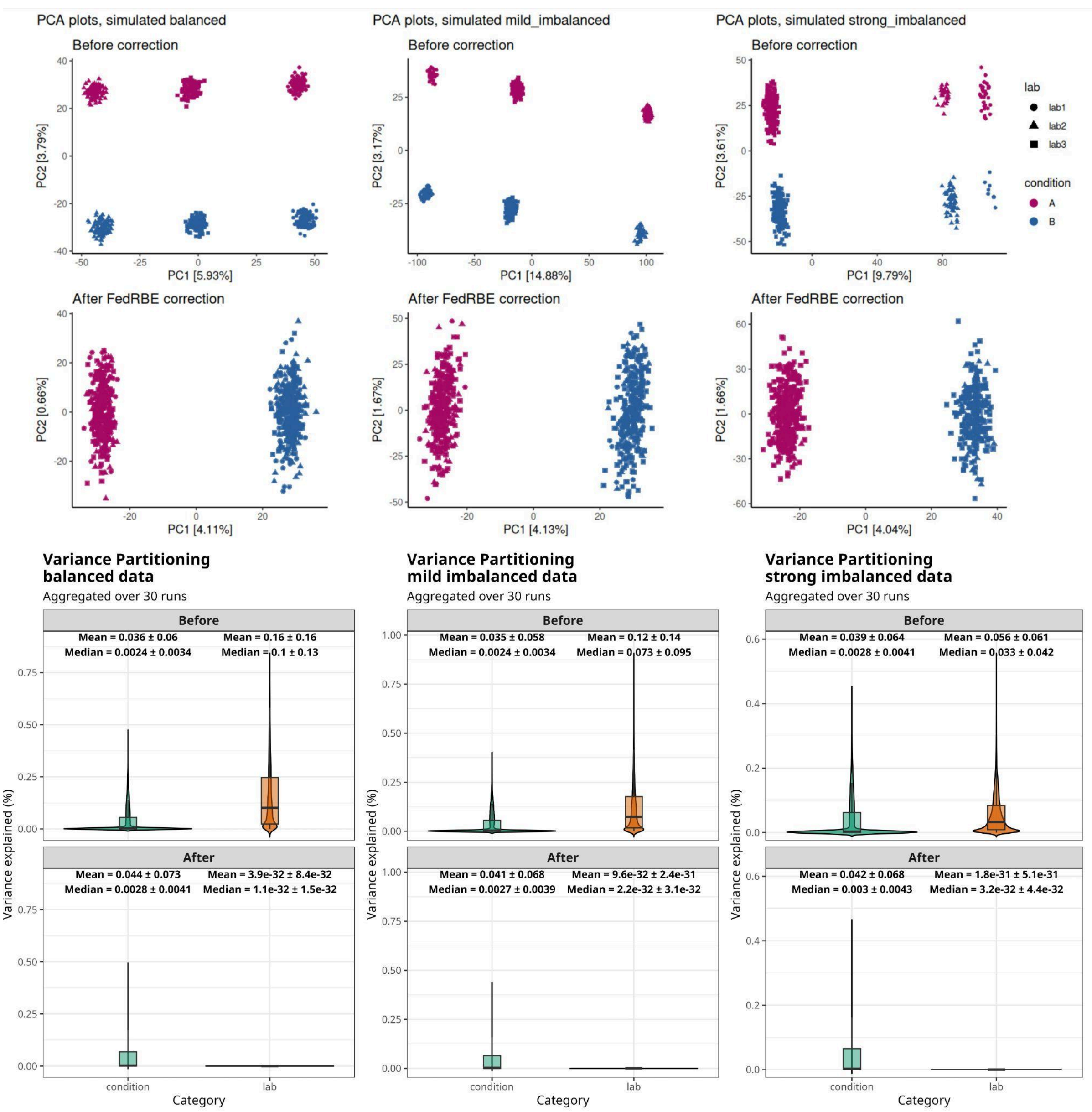

**Figure S4.** Results of batch effects correction with fedRBE on simulated data with different data imbalances. In panel a, the principal component analysis plots from one of the runs for each scenario are shown. For panel b, each simulation and analysis was repeated 30 times, and aggregated results were reported. The variance partitioning analysis results are shown for all 30 runs together. The mean and median are the mean and median of all runs separately, with the standard deviation indicated.

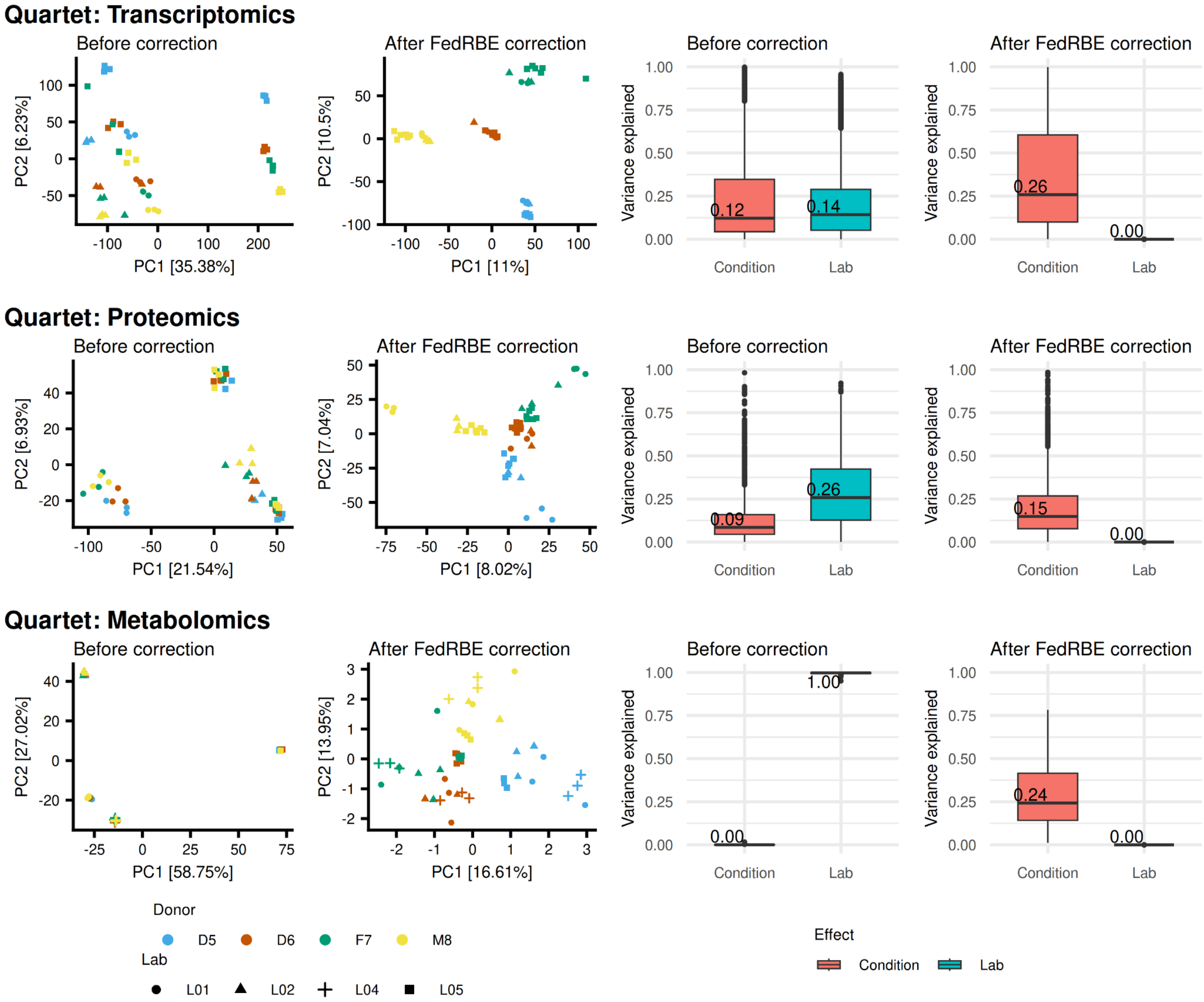

**Figure S5.** Separated modality PCA and fixed-effect variance-explained diagnostics for the Quartet multi-omics evaluation.

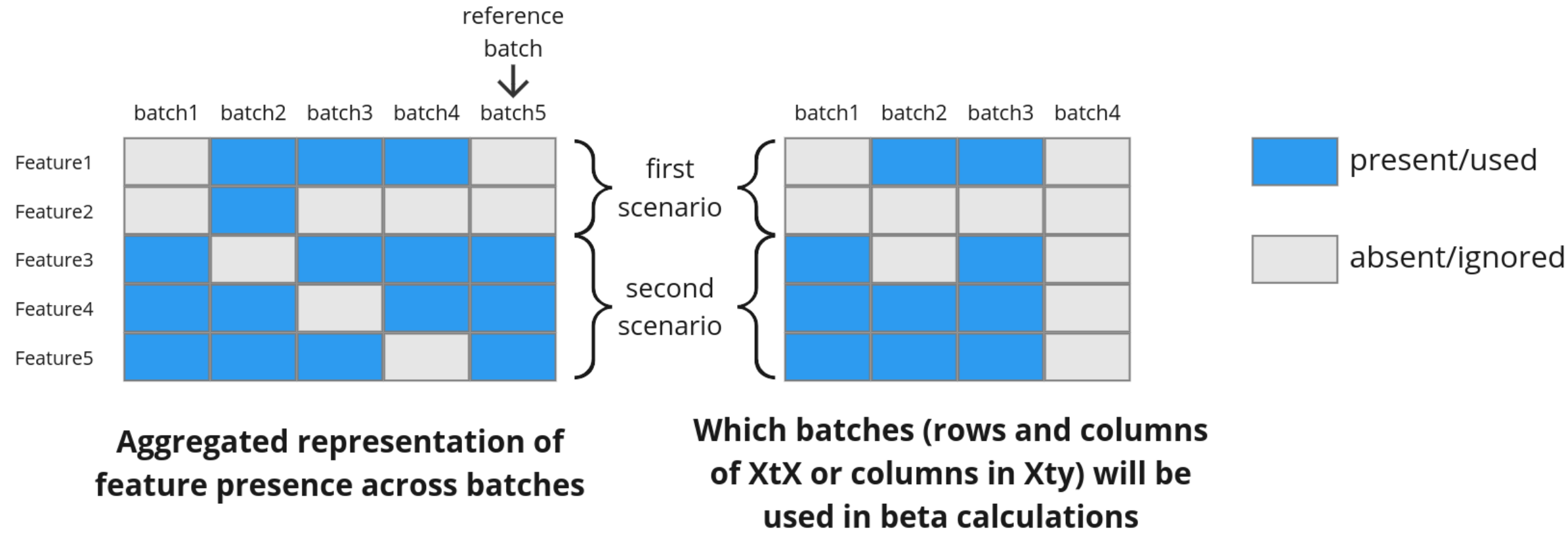

**Figure S6.** Global mask influence on beta calculations based on feature presence across batches. If the feature is absent in the reference batch, it is redefined by selecting another batch (last available) where the feature is present to serve as the new reference batch. If the feature is present in the reference batch but absent in any other, and if the last batch where it is present occurs after the first batch where it is absent, the positions of the batch effect terms are interchanged, corresponding to the first absent batch and the last present batch for feature $j$. In this way, the last present batch design parameters in $X$ are ignored, and the first absent batch data is used instead.

# Supplementary References